\documentclass[prd,aps,preprint,tightenlines,showkeys,showpacs,nofootinbib,superscriptaddress]{revtex4-1}
\usepackage{mathrsfs}
\usepackage{amsfonts}
\usepackage{amsmath}
 \usepackage{amssymb}
\usepackage{array}
\usepackage{verbatim}
\usepackage{bm}
\usepackage{epsfig}
\usepackage{graphicx}
\usepackage{relsize}
\RequirePackage{xspace}

\begin{document}
	\title{Photoproduction of vector mesons  in proton-proton ultraperipheral collisions at the Larger Hadron Collider}
	\author{Ya-Ping Xie}\email{xieyaping@impcas.ac.cn}
	\affiliation{Institute of Modern Physics, Chinese Academy of
		Sciences, Lanzhou 730000, China}
\affiliation{ University of Chinese Academy of Sciences, Beijing 100049, China}
	\author{Xurong Chen}\email{xchen@impcas.ac.cn}
	\affiliation{Institute of Modern Physics, Chinese Academy of
		Sciences, Lanzhou 730000, China}
	\affiliation{ University of Chinese Academy of Sciences, Beijing 100049, China}
	\begin{abstract}
		 Photoproduction of vector mesons are computed in dipole model in proton-proton
		 ultraperipheral collisions(UPCs)  at the CERN Larger Hadron Collider (LHC). The dipole model framework is employed in the calculations of cross sections of diffractive processes. Parameters of the bCGC model are refitted with the latest experimental data. The bCGC model and Boosted Gaussian wave functions are employed in the calculations. We obtain predictions of rapidity distributions of $J/\psi$ and $\psi(2s)$ mesons in proton-proton ultraperipheral collisions. The predictions give a good descriptions to the experimental data of LHCb. Predictions of $\phi$ and $\omega$ mesons are also calculated in this paper.
	\end{abstract}
	
	\maketitle
\section{introduction}
Diffractive photoproduction of vector mesons in hadron-hadron and electron-proton collisions can help us study the QCD dynamics and gluon saturation effect at high energy level \cite{Bertulani:2005ru,Baltz:2007kq}. The H1 and ZEUS collaborations have measured the cross sections of $J/\psi$ in diffractive process at HERA \cite{Chekanov:2002xi,Chekanov:2004mw,Aktas:2005xu,Alexa:2013xxa}. The LHCb collaborations have measured the rapidity distributions of $J/\psi$ and $\psi(2s)$ in proton-proton and nucleus-nucleus ultraperipheral collisions (UPCs) at the LHC\cite{Aaij:2013jxj,Aaij:2014iea,LHCb:2016oce,Abbas:2013oua,Abelev:2012ba,TheALICE:2014dwa,Adam:2015gsa,Adam:2015sia}.  Various theoretical approaches can be found to compute the production of vector mesons in UPCs and diffractive processes \cite{Klein:1999qj,Frankfurt:2002sv,Goncalves:2005yr,Ryskin:1992ui,Toll:2012mb,Adeluyi:2012ph,Xie:2016ino,Xie:2017mil}.\\
\indent In hadron-hadron UPCs, the direct hadronic interaction is suppressed. The photon-induced interaction is dominant in hadron-hadron UPCs. Vector mesons can be produced in photon-induced  process. The dipole model is a phenomenological model in small-x physics \cite{Forshaw:2003ki}. In the dipole model, the interaction between virtual photon and proton can be viewed as three steps.  Firstly, the virtual photon splits into quark and antiquark. Therefore, the quark-antiquark interacts with proton by exchange gluons. Finally, the quark-antiquark recombine into other particles, for example, vector mesons or real photon. The important aspect of dipole model is the cross section of a pair of quark-antiquark scattering off a proton via gluons exchange.  Dipole amplitude is the imaginary part of total photon-proton cross section. It is important in the diffractive process to calculate the production of vector mesons since the vector meson can be viewed as a probe of the interaction between the dipole and the proton.  The Golec-Biernat-Wusthoff (GBW) model was firstly introduced to describe the dipole cross section in saturation physics \cite{GolecBiernat:1998js}. The Bartel-Golec-Biermat-Kowalski (BGBK) model are extensive model of the GBW model considering the gluon density evolution according to DGLAP equation \cite{Bartels:2002cj}. The Color-Glass-Condensate (CGC) model was introduced based on Balitsky-Kovchegov (BK) evolution equation \cite{Iancu:2003ge,Soyez:2007kg,Ahmady:2016ujw}. The bSat and bCGC models are impact parameter dependent dipole models based on the BGBK and CGC models \cite{Kowalski:2003hm,Kowalski:2006hc,Rezaeian:2012ji,Watt:2007nr,Rezaeian:2013tka}. These models all contains free parameters which are determined by fit on cross sections of the inclusive production in deep inelastic scattering. \\
 \indent In the photoproduction of vector meson in diffractive process, the light-cone wave functions of photon and vector meson are employed in the amplitude. The light-cone wave function of photon can be analytically computed, but the light-cone function of the vector meson can't be computed analytically. The phenomenological models are used for the vector mesons. The Boosted Gaussian model is a successful model for $J/\psi$ and excited states.  The production of $J/\psi$ and $\psi(2s)$ can be used to check the validity of the Boosted Gaussian model.
Using the dipole amplitude and light-cone functions of photon and vector meson, the cross section in diffractive process can be evaluated as a function of Bjorken x. \\
\indent On other side, the cross sections of heavy vector mesons in diffractive process is investigated in perturbative QCD approach \cite{Jones:2013pga,Jones:2013eda,Jones:2016icr}. The vector meson amplitude is proportional to the gluon density.  The leptonic decay width of the heavy vector meson is included in the amplitude.  Rapidity gap survival factor is introduced in this paper \cite{Khoze:2013dha}. \\
\indent  In this paper, the  bCGC model is employed to perform the production of vector mesons in diffractive process. Then multiplying the photon flux and rapidity gap survival factor, we obtain the rapidity distributions of the vector mesons in proton-proton UPCs. Similar works can be found in Ref. \cite{Ducati:2013tva}. The aim of this paper is to update the prediction of exclusive production of $J/\psi$ and $\psi(2s)$ mesons and compute the rapidity distributions of $\phi$ and $\omega$ are performed in bCGC model using the Boosted Gaussian wave functions in proton-proton UPCs.  We obtain new parameters of bCGC model in this paper and we consider the 
contribution of rapidity  gap survival factors in this paper too.  In Section II, the theoretical framework is reviewed. In Section III, the parameters of the bCGC model are fitted using the latest experimental data. In section IV, the numerical results are presented and some discussions are also listed. The conclusions are in section IV.
\section{vector meson production in the dipole model}
In this paper, we focus on the production of heavy mesons in proton-proton UPCs. The rapidity distributions of heavy meson production in UPCs is the product of cross sections of $\gamma+\mathrm{p}\to V+\mathrm{p}$, the photon flux factor and rapidity gap survival factor. The rapidity distributions of heavy mesons in proton-proton UPCs is given as follows\cite{Jones:2013pga,Ducati:2013tva}
\begin{eqnarray}
\frac{d\sigma}{dy}=S^2(W^+)k^+\frac{dn}{dk^+}(k^+)\sigma^{\gamma p\to Vp}(W^+)+S^2(W^-)k^-\frac{dn}{dk^-}(k^-)\sigma^{\gamma p\to Vp}(W^-).
\label{dsdy}
\end{eqnarray}
In above equation, $k$ is momentum of the radiated photon from proton. $\mathrm{y}$ is the rapidity of the vector meson.  $k^{\pm}=M_V/2\exp(\pm |\mathrm{y}|)$.  $\mathrm{W}^{\pm}$ is the center mass energy in diffractive process In UPCs, $W^{\pm}=(2k^\pm\sqrt{s})^{1/2}$ with $\sqrt{s}$ center-energy.   $S^2(W)$ is rapidity gap survival factor in Good-Walker model\cite{Jones:2013pga,Khoze:2002dc}, and $dn/dk$ is photon flux\cite{Bertulani:2005ru}. It is given by
\begin{eqnarray}
\frac{dn}{dk}(k)=\frac{\alpha_{em}}{2\pi k}\Big[1+\Big(1-\frac{2k}{\sqrt{s}}\Big)^2\Big]\Big(
\ln \Omega-\frac{11}{6}+\frac{3}{\Omega}-\frac{3}{2\Omega^2}+\frac{1}{3\Omega^3}\Big),
\end{eqnarray}
where $\Omega=1+0.71/Q^2_{min}$, with $Q^2_{min}=k^2/\gamma^2_L$, $\gamma_L$ is the lorentz boost factor with $\gamma_L=\sqrt{s}/2m_p$.
The cross sections of $\sigma^{\gamma p\to Vp}(W)$ is integrated by $|t|$ as.
\begin{equation}
\sigma^{\gamma p\to Vp}(W)=\int dt\frac{d\sigma^{\gamma p\to Vp}}{dt}.
\end{equation}
Then, the differential cross section of $\gamma+p\to V+p$ is given as \cite{Kowalski:2003hm,Kowalski:2006hc}
\begin{eqnarray}
\frac{d\sigma^{\gamma p\to Vp}(x)}{dt}=\frac{R_g^2(1+\beta^2)}{16\pi^2}
|\mathcal{A}^{\gamma p\to Vp}(x,Q^2,\Delta)|^2,
\label{dsigma1}
\end{eqnarray}
with $x=\frac{M_V}{\sqrt{s}}\exp(\mp| \mathrm{y|})$ or $x=M_V^2/W^2$. The amplitude $\mathcal{A}^{\gamma p\to Vp}(x,Q^2,\Delta)$ in Eq.~(\ref{dsigma1}) is written as
\begin{eqnarray}
\mathcal{A}^{\gamma p\to Vp}(x, Q^2,\Delta)= i\int
d^2r\int_0^1\frac{dz}{4\pi} \int
d^2b(\psi_V^*\psi_{\gamma})_{T}(z,r,Q^2)e^{-i(\bm b-(1-z)\bm r)\cdot\bm
	\Delta }\mathcal{N}(x,\bm r,\bm b),\notag\\
\label{amp}
\end{eqnarray}
where T denotes the transverse overlap function of photon and vector meson functions with $Q^2=0$, since the photon is real one in UPCs.  And $\beta$ is ratio of the imaginary part to the real part amplitude.
\begin{equation}
\beta=\tan (\frac{\pi}{2}\delta),  \quad\text{with}\quad \delta=\frac{\partial \ln (\mathrm{Im}\mathcal{A}(x))}{\partial \ln(1/x)}.
\end{equation}
The factor $R_g^2$ reflects the skewedness \cite{Shuvaev:1999ce}, it gives
\begin{equation}
R_g=\frac{2^{2\delta+3}}{\sqrt{\pi}}\frac{\Gamma(\delta+5/2)}{\Gamma(\delta+4)}.
\end{equation}
  In the bCGC model, the dipole amplitude is given as \cite{Iancu:2003ge,Rezaeian:2013tka}
  \begin{eqnarray}
\mathcal{N}(x,\bm r,\bm b)=2\times\begin{cases}
  \mathcal{N}_0(\frac{rQs}{2})^{2(\gamma_s+(1/\kappa\lambda Y)\ln(2/rQs))},\quad\! rQs\leqslant 2,\\
  1-\exp\big(-A\ln^2(B rQs)\big),\quad\quad\!\! rQs>2,
  \end{cases}
  \end{eqnarray}
  where $Qs(x,\bm b)=(x/x_0)^{\lambda/2}\exp(-\frac{\bm b^2}{4\gamma_sB_p})$, $\kappa=9.9$, and $Y=\ln(1/x)$. $A$ and $B$ are given as
   \begin{equation}
   \begin{split}
   & A=-\frac{\mathcal{N}^2_0\gamma_s^2}{(1-\mathcal{N}_0)^2\ln(1-\mathcal{N}_0)},\\
   &B=\frac{1}{2}(1-\mathcal{N}_0)^{-(1-\mathcal{N}_0)/(2\mathcal{N}_0\gamma_s)}.
   \end{split}
   \end{equation}
   In the bCGC model, $B_p$, $x_0$, $\gamma_s$, $\mathcal{N}_0$ and $\lambda$ are free parameters and they are fitted from the experimental data.\\
  \indent The overlap of photon and vector meson in Eq.~(\ref{amp}) we use are given as follows
   \begin{eqnarray}
   (\Psi_V^*\Psi_{\gamma})_T(r,z,Q^2)=e_fe\frac{N_c}{\pi z(1-z)}\lbrace  m_f^2
   K_0(\epsilon r)\phi_T(r,z)-(z^2+(1-z)^2)\epsilon K_1(\epsilon r)\partial_r
   \phi_T(r,z)\rbrace,\notag\\
   \end{eqnarray}
   where $e_f$ is effective  charge for mesons, $\epsilon=\sqrt{z(1-z)Q^2+m_f^2}$ and $\phi_T(r,z)$ is the scalar functions, $K_0(x)$ and $K_1(x)$ are second kind Bessel functions. There is no analytic expression for the scalar functions of the vector mesons. There are some successful models for the scalar functions.
   The Boosted Gaussian model is a phenomenological model. The scalar function of $J/\psi$ in Boosted Gaussian model is written as
   \begin{eqnarray}
   \phi^{1s}_T(r,z)=N_Tz(1-z)\exp\big(-\frac{m_f^2\mathcal{R}_{1s}^2}{8z(1-z)}-
   \frac{2z(1-z)r^2}{\mathcal{R}^2_{1s}}+\frac{m_f^2\mathcal{R}^2_{1s}}{2}\big).
   \end{eqnarray}
   The scalar function for $\psi(2s)$ meson in Boosted Gaussian model is given as~\cite{Armesto:2014sma}
   \begin{eqnarray}
   \phi^{2s}_T(r,z)&=&N_Tz(1-z)\exp\big(-\frac{m_f^2\mathcal{R}^2_{2s}}{8z(1-z)}-
   \frac{2z(1-z)r^2}{\mathcal{R}^2_{2s}}+\frac{m_f^2\mathcal{R}^2_{2s}}{2}\big)\notag\\
   &\times&\Big[1+\alpha_{2s}\Big(2+\frac{m_f^2\mathcal{R}^2_{2s}}{8z(1-z)}-
   \frac{4z(1-z)r^2}{\mathcal{R}_{2s}^2}-m_f^2\mathcal{R}_{2s}^2\Big)\Big].
   \end{eqnarray}
There are several free parameters of the Boosted Gaussian wave functions. They are presented in Table. \ref{wave}.  The parameters of $\omega$ meson are obtained in this work.
The parameters are determined by the normalization condition and the lepton decay width.
\begin{table}[htbp]
\begin{tabular}{p{1cm} | p{1.5cm}p{1.5cm}p{1.5cm}p{1.5cm}p{1.5cm}p{1.5cm} p{1.5cm}}
\hline
\hline
meson &$e_f$& mass& $f_V$ & $m_f$& $N_T$& $\mathcal{R}^2$&$\alpha_{2s}$ \\
\hline
	& & GeV & GeV & GeV &   & $\text{GeV}^2 $ & \\
\hline
    $\omega$  &$1/3\sqrt{2}$      & 0.782    & 0.0458  & 0.14      & 0.895     &15.78 & \\
      $\phi$  &$1/3$      & 1.020    & 0.076  & 0.14    &   0.919    &11.2 & \\
    $J/\psi$  &$2/3$      & 3.097    & 0.274  & 1.27      & 0.596      &2.45 & \\
$J/\psi$  &$2/3$      & 3.097    & 0.274  & 1.40   & 0.57     &2.45 & \\
$\psi(2s)$  &$2/3$      & 3.686  & 0.198  & 1.27      & 0.70      &3.72 & -0.61\\
$\psi(2s)$  &$2/3$      & 3.686  & 0.198  & 1.40      & 0.67      &3.72 & -0.61\\
\hline
\hline
\end{tabular}
\caption{Parameters of the scalar functions of the Boosted Gaussian model for $\omega$, $\phi$, $J/\psi$ and $\psi(2s)$ mesons, the parameters of $\rho$, $J/\psi$ and $\psi(2s)$ are taken from~\cite{Kowalski:2006hc,Armesto:2014sma}.}
\label{wave}
       \end{table}
       \section{parameters fit for the bCGC model}
In the bCGC model, there are several free parameters need to be fitted from the experimental data. In dipole model, the cross section of the virtual photon and the proton in Deep Inelastic Scattering (DIS) are written as
     \begin{eqnarray}
     \sigma_{T,L}^{\gamma^*p}(x,Q^2)=&&\sum_{f=u,d,s}\int d^2 \bm r\int \frac{dz}{4\pi}(\psi^*\psi)_{T,L}^f(z,\bm r, Q^2)
     \sigma_{q\bar{q}}(x,\bm r)\notag\\
     &&+\sum_{f=c}\int d^2 r\int \frac{dz}{4\pi}(\psi^*\psi)_{T,L}^f(z,\bm r,Q^2)
     \sigma_{q\bar{q}}(\hat{x},\bm r).
     \label{dipolecross}
     \end{eqnarray}
       where $x=x_{Bjorken}$ and $\hat{x}=x_{Bjorken}(1+4m^2_c/Q^2)$. The dipole cross section $\sigma_{q\bar{q}}(x,\bm r)$ is integrated as
       \begin{eqnarray}
       \sigma_{q\bar{q}}(x,\bm r)=\int d^2\bm b\mathcal{N}(x,\bm r,\bm b).
       \end{eqnarray}
       The square of the wave functions of the virtual photons are given by
          \begin{eqnarray}
          (\psi^*\psi)_T(z,\bm r,Q^2)&=&\frac{2N_c}{\pi}\alpha_{em}e_f^2\{[z^2+(1-z)^2]\epsilon^2K^2_1(\epsilon r)
          +m_f^2K^2_0(\epsilon r)\}; \\
          (\psi^*\psi)_L(z,\bm r, Q^2)&=&\frac{8N_c}{\pi}\alpha_{em}e_f^2Q^2z^2(1-z)^2K_0^2(\epsilon r).
          \end{eqnarray}
          The proton structure functions $F_2(x,Q^2)$ and $F_L(x,Q^2)$ are written as
         	\begin{eqnarray}
	F_2(x,Q^2)&=&\frac{Q^2}{4\pi^2\alpha_{em}}[\sigma_T^{\gamma^*p}(x,Q^2)+\sigma_L^{\gamma^*p}(x,Q^2)];\\
	F_L(x,Q^2)&=&\frac{Q^2}{4\pi^2\alpha_{em}}\sigma^{\gamma^*p}_L(x,Q^2).
	\end{eqnarray}
	The reduce cross section in DIS is given by
\begin{eqnarray}
\sigma_r(x,\text{y},Q^2)=F_2(x,Q^2)-\frac{\text{y}^2}{1+(1-\text{y})^2}F_L(x,Q^2).
\end{eqnarray}
where $\text{y}=Q^2/(xs)$. In 2015, H1 and ZEUS released the latest combined reduce cross sections \cite{Abramowicz:2015mha}. \\
\indent In this paper, we refit the free parameters of the bCGC model using the reduce cross sections released in 2015. The experimental data are selected from $x<0.01$ and $0.40\;\text{GeV}^2\leqslant Q^2\leqslant 45\;\text{GeV}^2$. The parameters fitted in this paper are presented in Table~\ref{IPP2} with two fits.
\begin{table}[h]
	\begin{tabular}{p{1cm}p{2.0cm}p{1.5cm}p{1.8cm}p{1.2cm}p{1.2cm}p{1.6cm}p{1.25cm}p{2.5cm}}
		\hline
		\hline
		&$m_{u,d,s}$/GeV&$m_{c}$/GeV&	$B_p$/~$\mathrm{GeV}^2$ & $\gamma_s$ & $N_0$ & $x_0$ & $\lambda$ & $\chi^2$/d.o.f\\
		\hline
		Fit 1&  0.14&1.27&	5.746& 0.6924    & 0.3159&
		0.001849&  0.2039  &  607/467=1.300\\
		Fit 2&0.14&1.4    &  	5.852  &0.6932 &0.3144  &
		0.001978&     0.2012 & 629/467=1.347 \\
		\hline
		\hline
	\end{tabular}
	\caption{Parameters for bCGC model fitted from the reduce cross sections with $x<0.01$ and $0.40\; \text{GeV}^2\leqslant Q^2\leqslant45\;\text{GeV}^2$ released in 2015 \cite{Abramowicz:2015mha}.}
	\label{IPP2}
\end{table}

       \begin{figure}[htbp]
       \centering
       \includegraphics[width=5in]{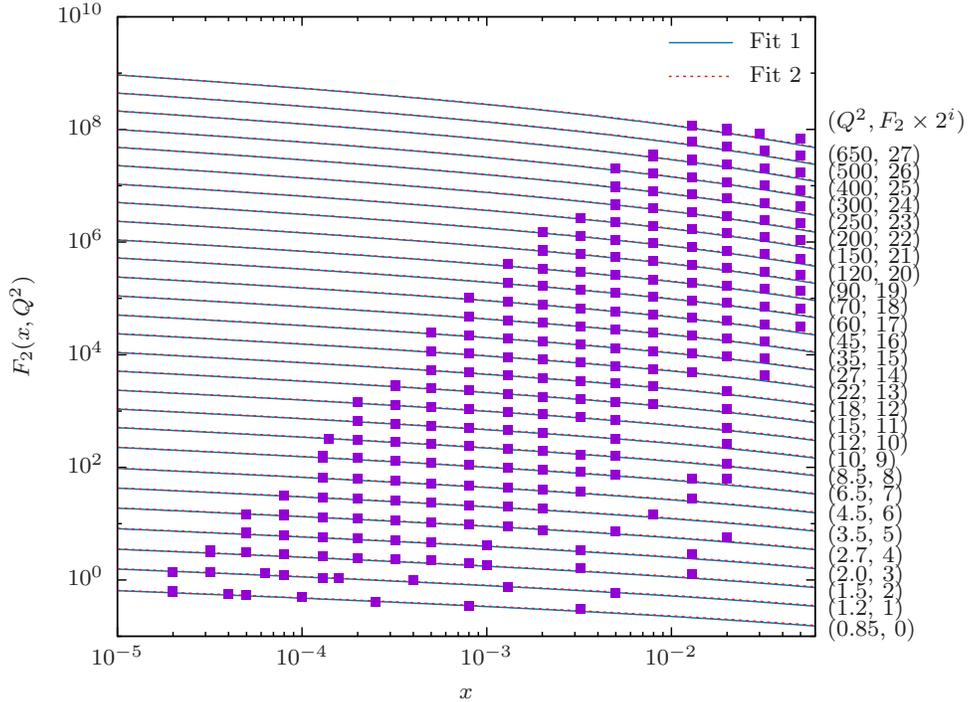}
       \caption{(Color online)  Proton structure functions $F_2(x,Q^2)$ calculated in bCGC model with parameters presented in Table~\ref{IPP2} and compared with the experimental data from H1 and ZEUS collaboration \cite{Aaron:2009aa}.}
       \label{F2}
    \end{figure}
  \begin{figure}[htbp]
  	\centering
  	\includegraphics[width=4in]{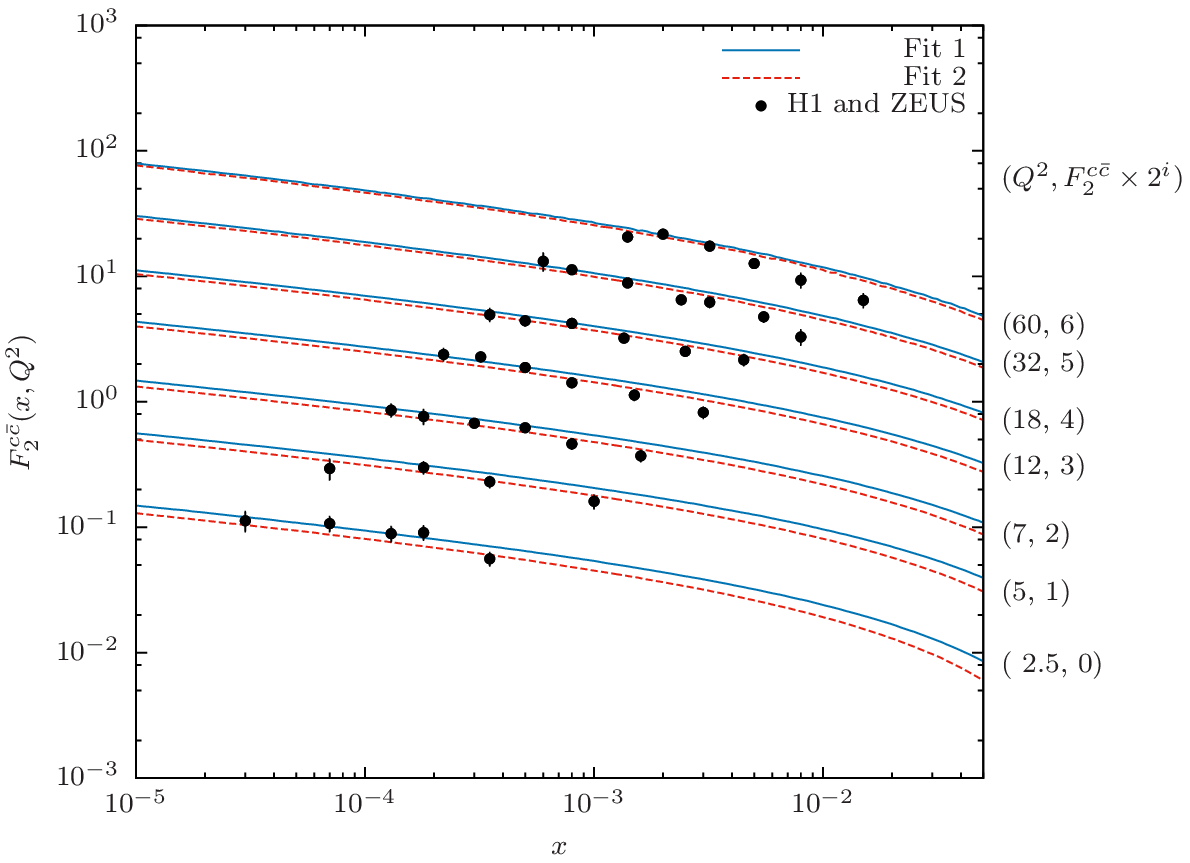}
  	\caption{(Color online)   Proton charmed structure functions $F^{c\bar{c}}_2(x,Q^2)$ calculated in bCGC model with parameters presented in Table~\ref{IPP2} and compared with charmed structure functions $F^{c\bar{c}}_2\approx \sigma_r^{c\bar{c}}$ from H1 and ZEUS collaboration \cite{Abramowicz:1900rp}}
  	\label{FCC}
  \end{figure}
Using the parameters of the bCGC model, the proton structure function $F_2(x,Q^2)$ can be evaluated in the bCGC model and compared with experimental data. The proton structure functions for proton is shown in Fig.~\ref{F2}. It can be seen that the bCGC model give a good description to structure function $F_2(x,Q^2)$ using two fits parameters. The charmed proton structure function $F^{c\bar{c}}_2(x,Q^2)$ is shown In Fig.~\ref{FCC}, it can be seen that the two fits parameters give different predictions for the charmed structure function. 
 \section{numerical results and discussions}
\begin{figure}[htbp]
     	\centering
     	\includegraphics[width=4in]{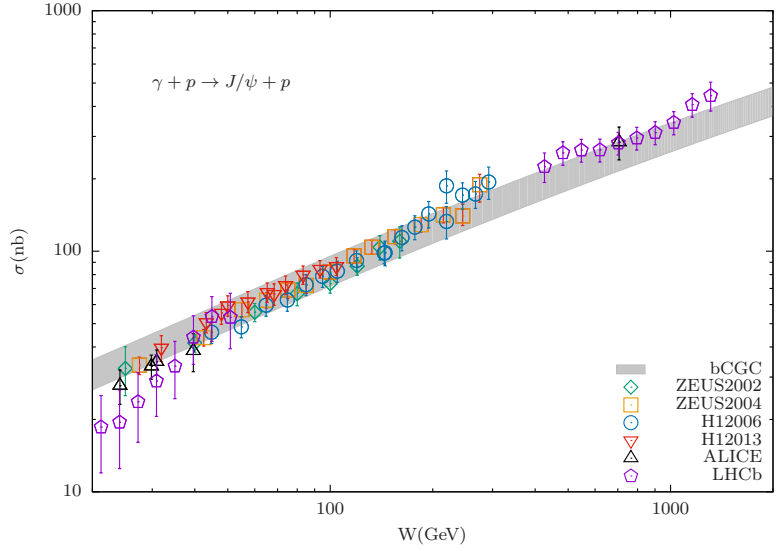}
     	\caption{(Color online)  Predictions of cross sections of diffractive process as a function of W calculated in the bSat and bCGC models with the Boosted Gaussian wave functions compared with the experimental data from H1\cite{Aktas:2005xu,Alexa:2013xxa}, ZEUS \cite{Chekanov:2002xi,Chekanov:2004mw}, ALICE \cite{TheALICE:2014dwa} and LHCb \cite{Aaij:2014iea}, The upper band of bCGC are using parameters with $m_c=1.27$ GeV and the lower band of bCGC are using parameters with $m_c=1.4$ GeV.}
     	\label{sigma}
     \end{figure}

\begin{figure}[!h]
\centering
\includegraphics[width=3in]{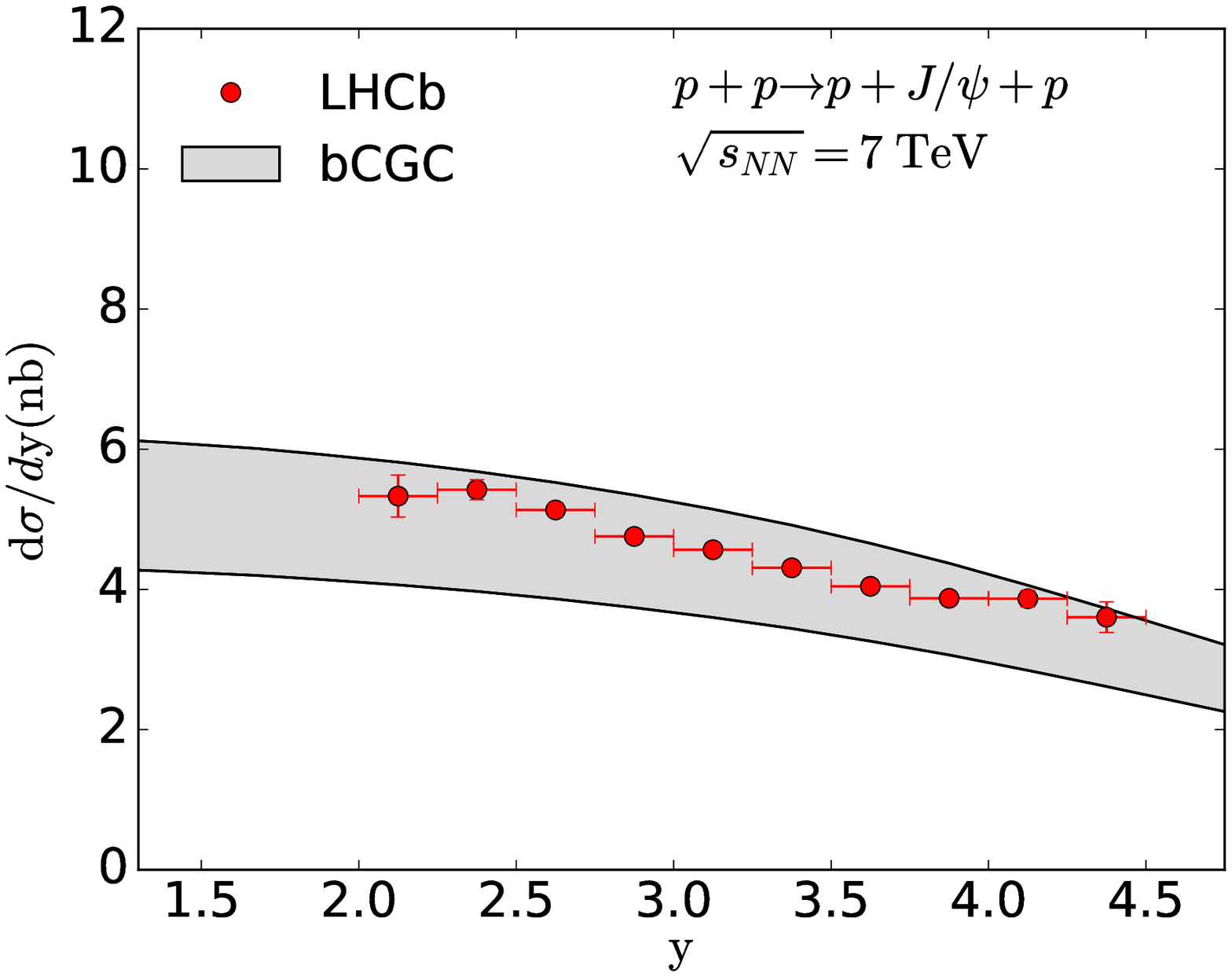}
\includegraphics[width=3in]{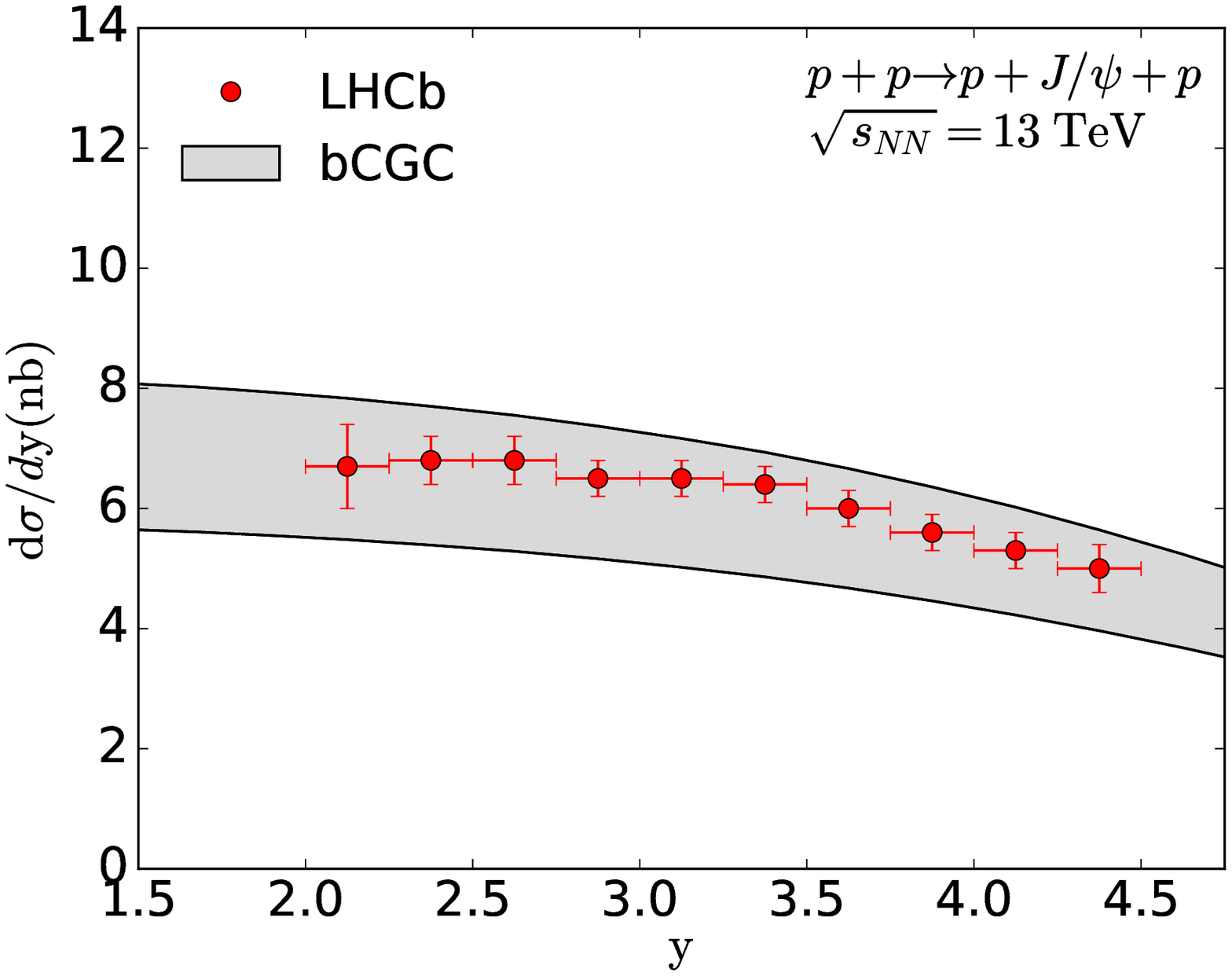}
\caption{(Color online) Predictions of rapidity distributions of $J/\psi$ meson in proton-proton ultraperipheral collisions at the LHC compared with the experimental data of the LHCb collaboration\cite{Aaij:2014iea,LHCb:2016oce}. The upper band of bCGC are using parameters with $m_c=1.27$ GeV and the lower band of bCGC are using parameters with $m_c=1.4$ GeV.}
	\label{jpsi}
\end{figure}

\begin{figure}[!h]
\centering
\includegraphics[width=3in]{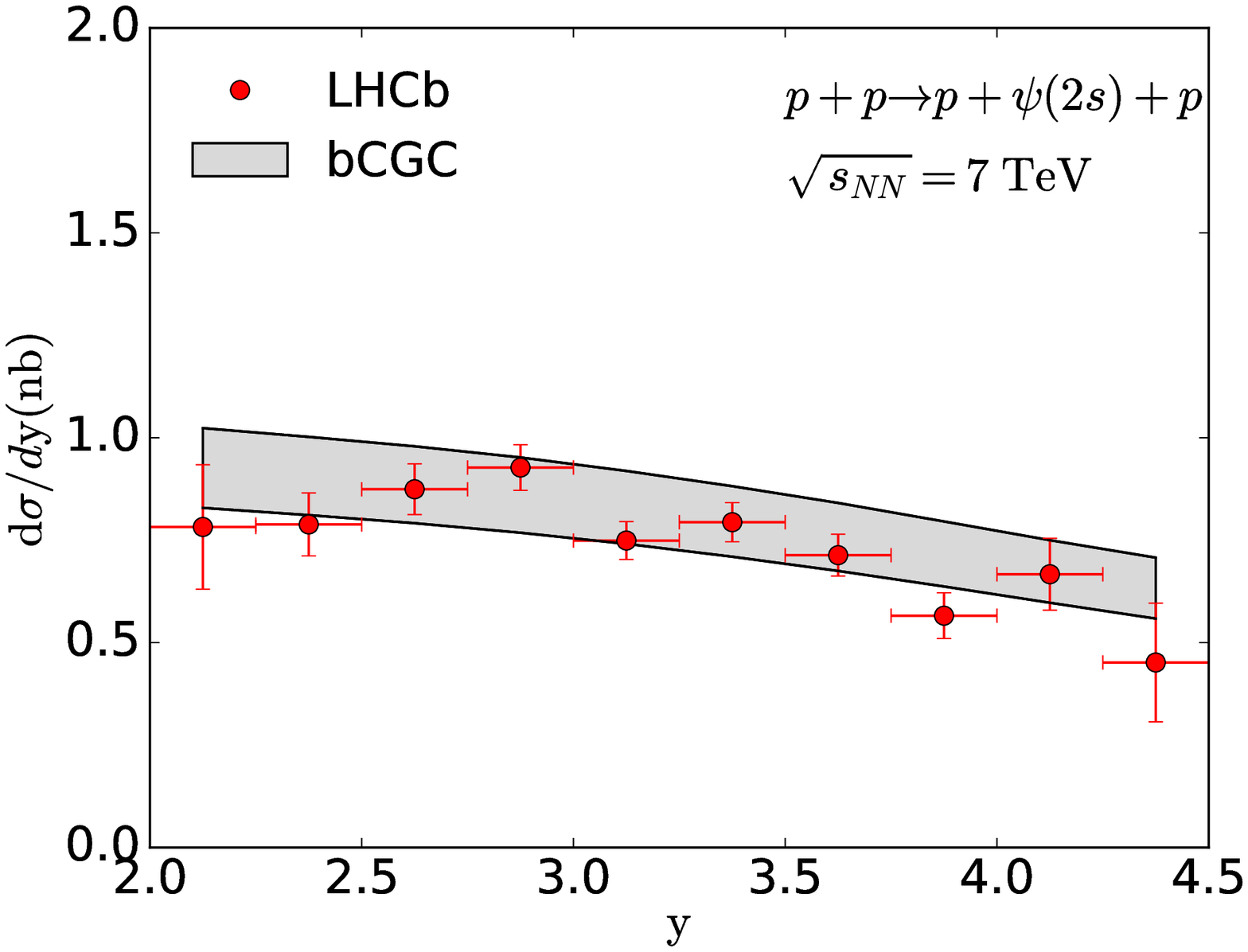}
 \includegraphics[width=3in]{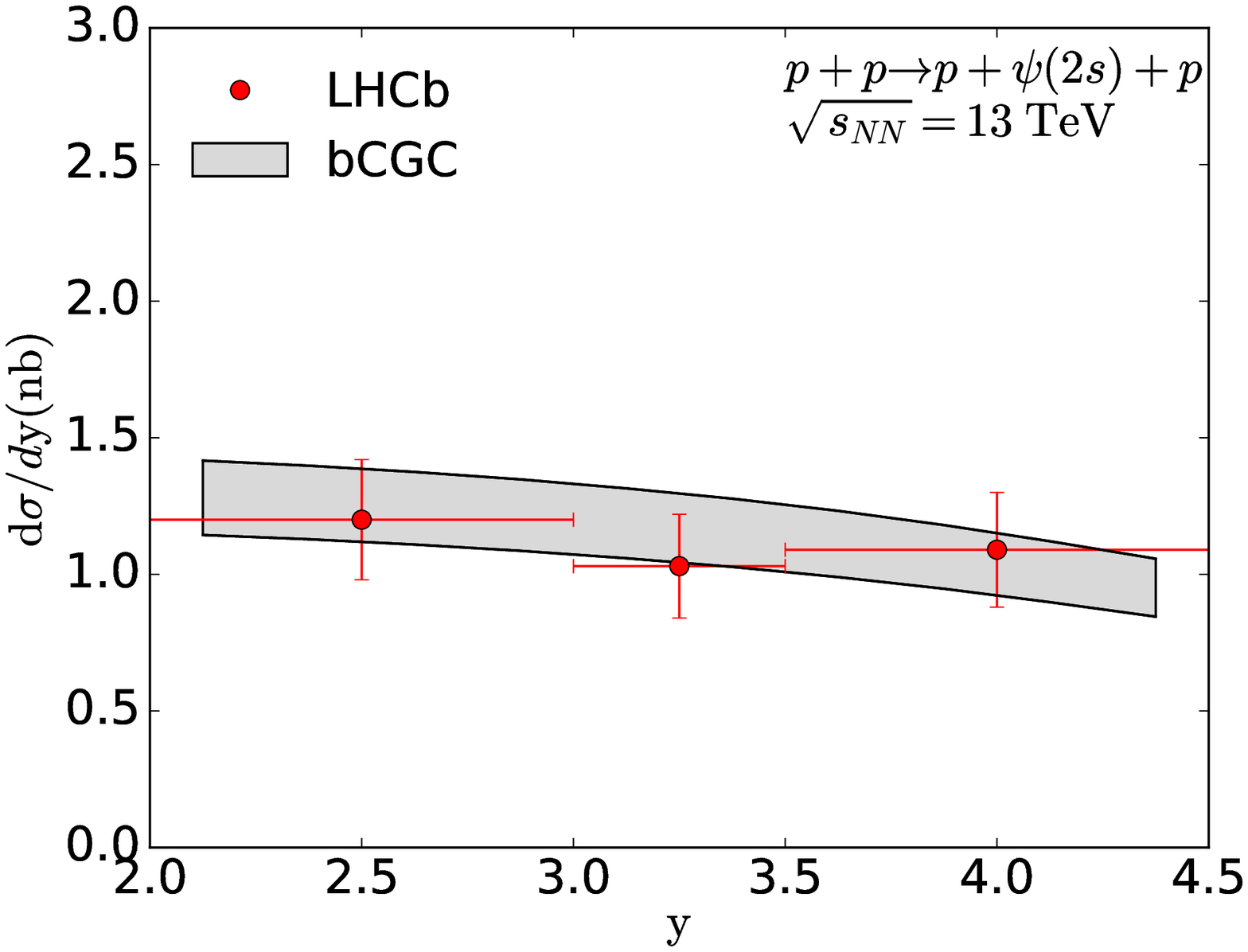}
\caption{(Color online) Predictions of rapidity distributions of $\psi(2s)$ meson in proton-proton ultraperipheral collisions at the LHC compared with the experimental data of the LHCb collaboration\cite{Aaij:2014iea,LHCb:2016oce}. The upper band of bCGC are using parameters with $m_c=1.27$ GeV and the lower band of bCGC are using parameters with $m_c=1.4$ GeV.}
\label{psi2s}
\end{figure}

Firstly, we compute the cross sections of $J/\psi$ in diffractive process and compare the predictions with experimental data. The amplitude of $\gamma+ p\to J/\psi +p$ are performed in bCGC model with the Boosted Gaussian wave functions. The parameters with $m_c=1.27$ GeV and $m_c=1.4$ GeV  in Table. \ref{IPP2} are used in the calculations. Predictions of cross section of $J/\psi$ meson in diffractive process are shown in Fig. \ref{sigma}. The upper band of bCGC are using parameters with $m_c=1.27$ GeV and the lower band of bCGC are using parameters with $m_c=1.4$ GeV as presented in Table \ref{IPP2}. It can be seen that the predictions using parameters $m_c$=1.27 GeV give a better description than the fit with $m_c$=1.4 GeV.  It can be concluded that the dipole model is sensitive to the quark mass. The cross sections labeled H1 and ZEUS are measured directly by H1 and ZEUS collaboration. The cross section labeled ALICE and LHCb are also not measured directly. They are extracted from p-Pb and proton-proton UPCs. The cross sections of LHCb are divided by the rapidity gap survival factor and photon flux as presented Eq.~(\ref{dsdy}). Therefore, we need add the rapidity gap survival factor contribution as Eq.~(\ref{dsdy}). The rapidity gap survival factor we use are from Refs. \cite{LHCb:2016oce,Jones:2016icr}.\\
 \indent  Secondly, we compute the rapidity distributions of $J/\psi$ and $\psi(2s)$ mesons as Eq. (\ref{dsdy}). The rapidity gap survival factors and photon flux are included. The parameters in Table. \ref{IPP2} of bCGC model are used in the calculations. The rapidity distributions of $J/\psi$ and $\psi(2s)$  mesons computed in two fits parameters are shown in Fig.~\ref{jpsi} and Fig.~\ref{psi2s} . The experimental data of LHCb are also presented in the same figures. The upper band of bCGC are using parameters with $m_c=1.27$ GeV and the lower band of bCGC are using parameters with $m_c=1.4$ GeV. It can be seen that our predictions give a good descriptions to the experimental data.  In Ref. \cite{Ducati:2013tva}, the rapidity distributions of $J/\psi$ and $\psi(2s)$ mesons had been computed in CGC model using the Boosted Gaussian wave functions, but the parameters of the Boosted Gaussian functions were not presented in Ref. \cite{Ducati:2013tva}. The predictions of this paper are close to the results in Ref. \cite{Ducati:2013tva,Fiore:2014oha}. We use the bCGC models with parameters fitted from combined H1 and ZEUS data and we present the detail parameters for the Boosted Gaussian wave functions and rapidity gap survival factors.  In Ref. \cite{Goncalves:2016sqy}, the rapidity distributions $J/\psi$ are obtained in the bCGC model, but the rapidity gap survival factor is unity. In our calculation, the rapidity gap survival factor is about $0.6\sim 0.9$. And we find that the rapidity gap survival factor is important in the final results of rapidity distributions. \\

 \begin{figure}[!h]
 	\centering
 	\includegraphics[width=3in]{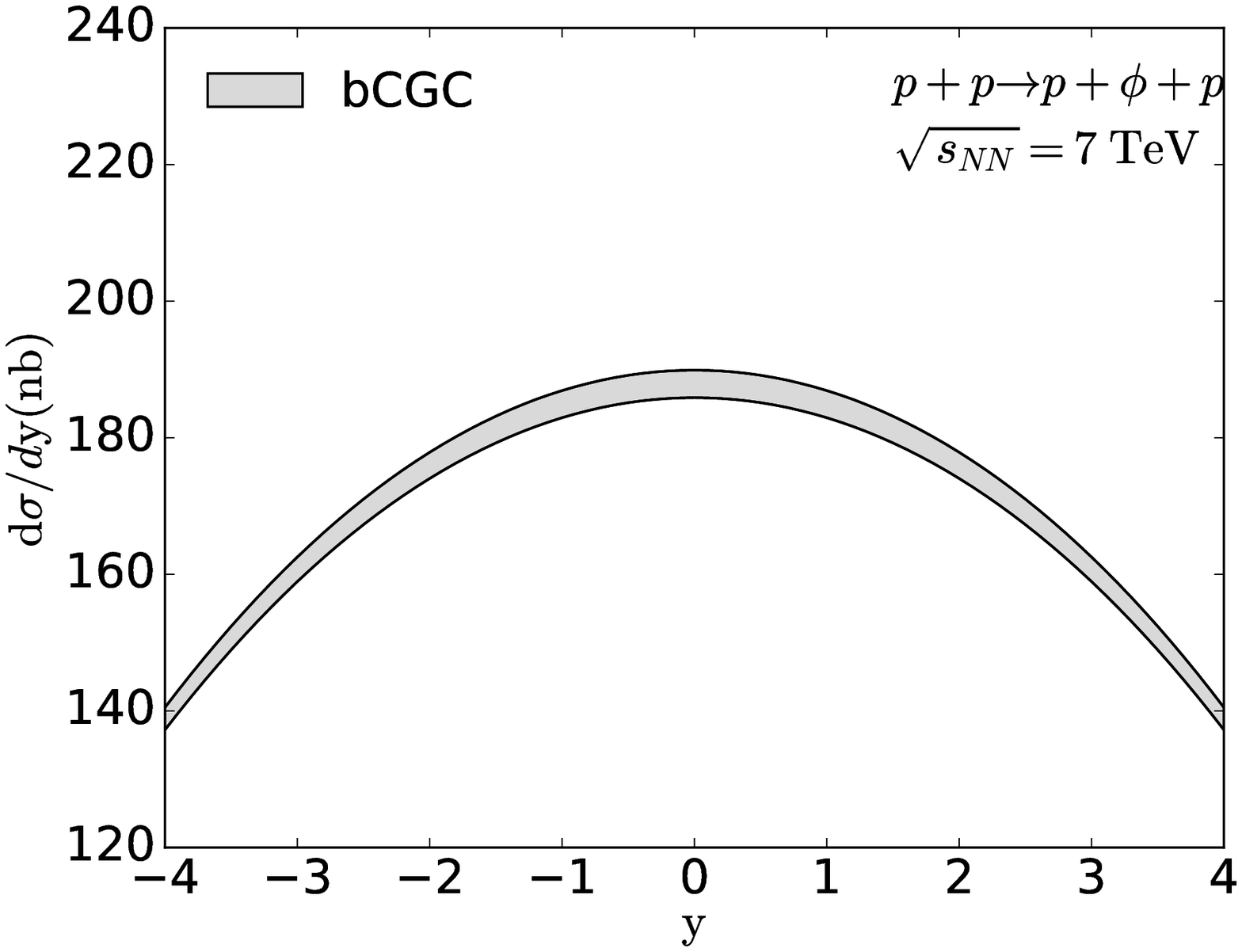}
 	\includegraphics[width=3in]{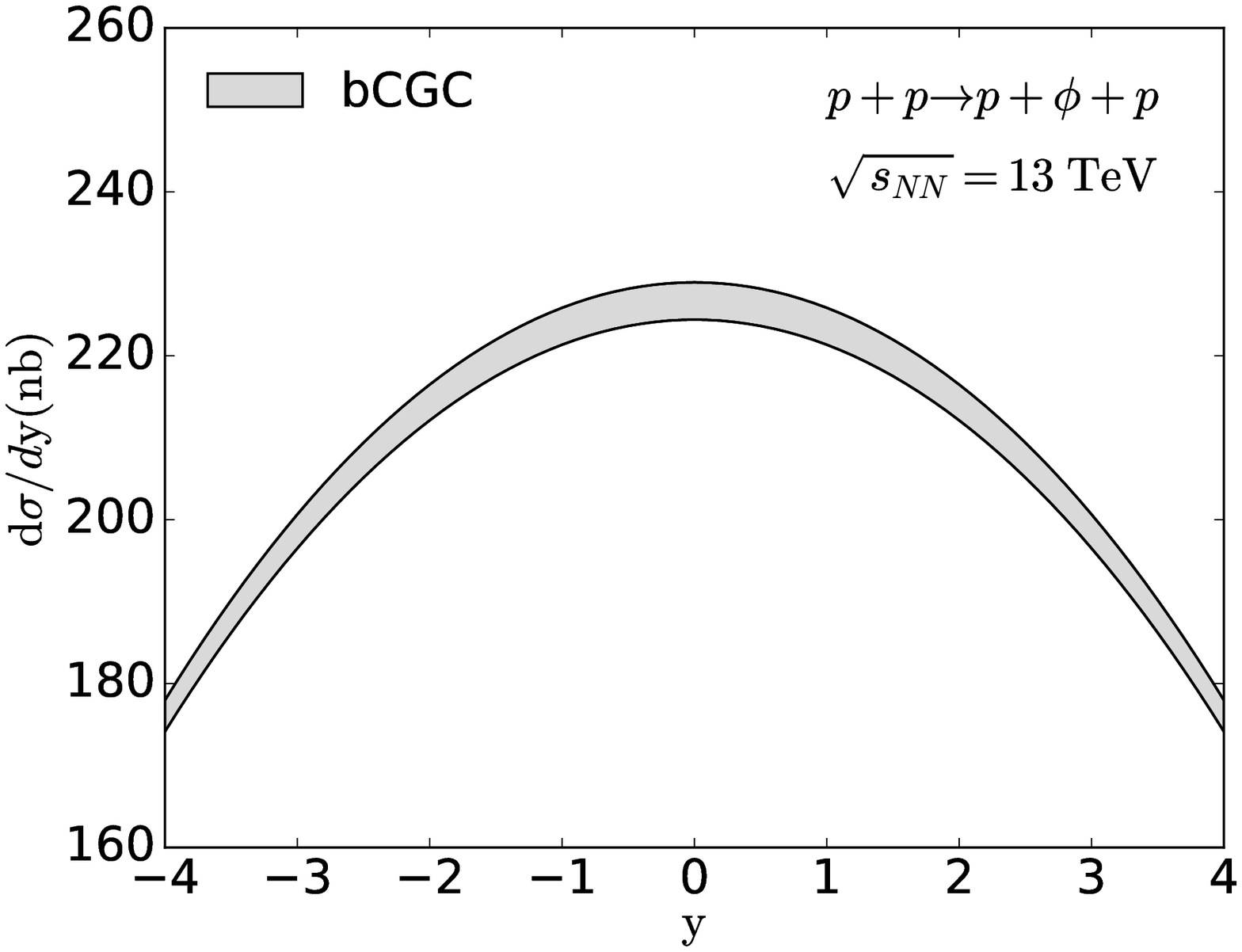}
 	\caption{(Color online) Predictions of rapidity distributions of $\phi$  meson computed in bCGC model using the Boosted Gaussian wave function in proton-proton ultraperipheral collisions at the LHC. The upper band of bCGC are using parameters of Fit 2 and the lower band of bCGC are using parameters of Fit 1.}
 	\label{phi}
 \end{figure}
  \begin{figure}[!h]
  	\centering
  	\includegraphics[width=3in]{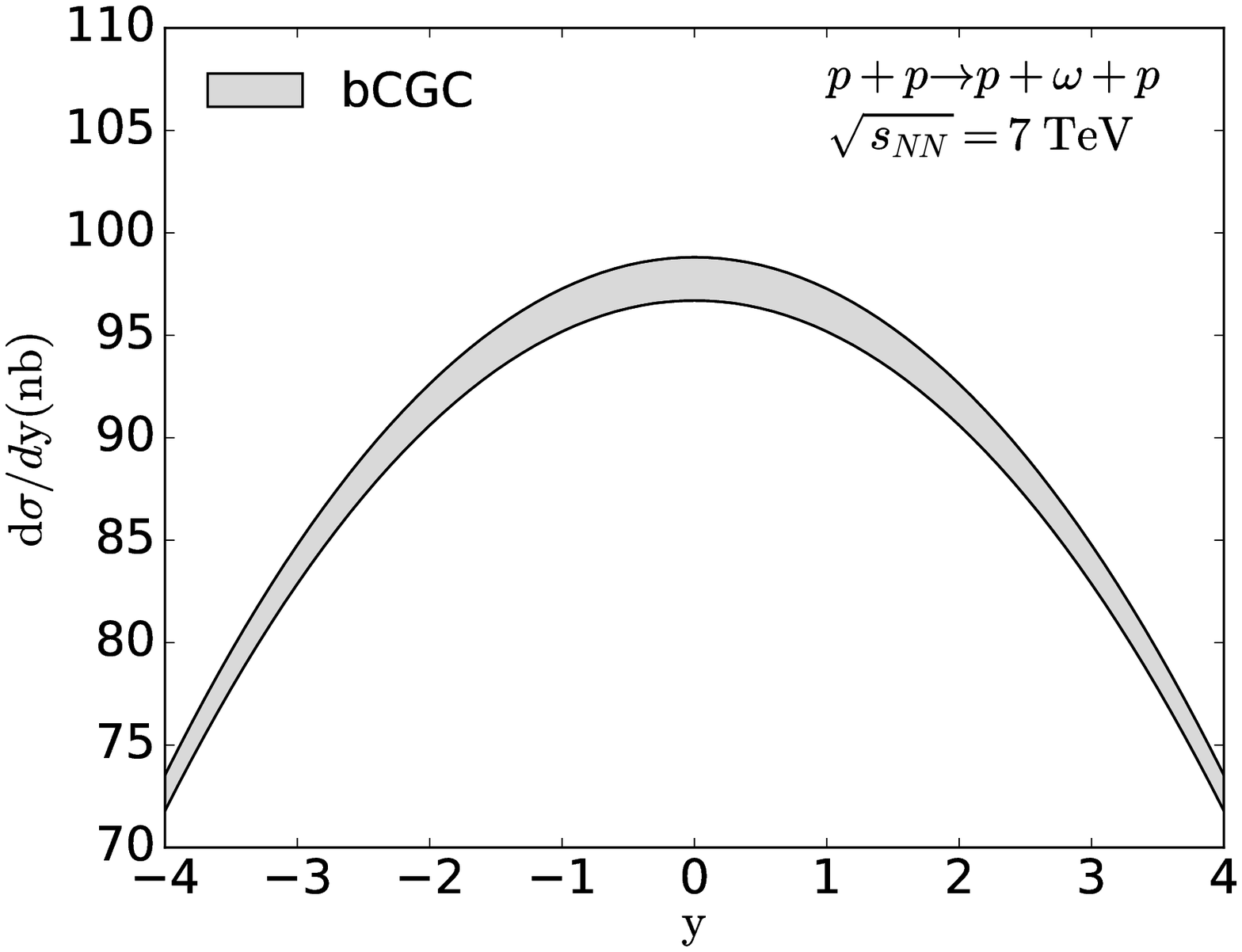}
  	\includegraphics[width=3in]{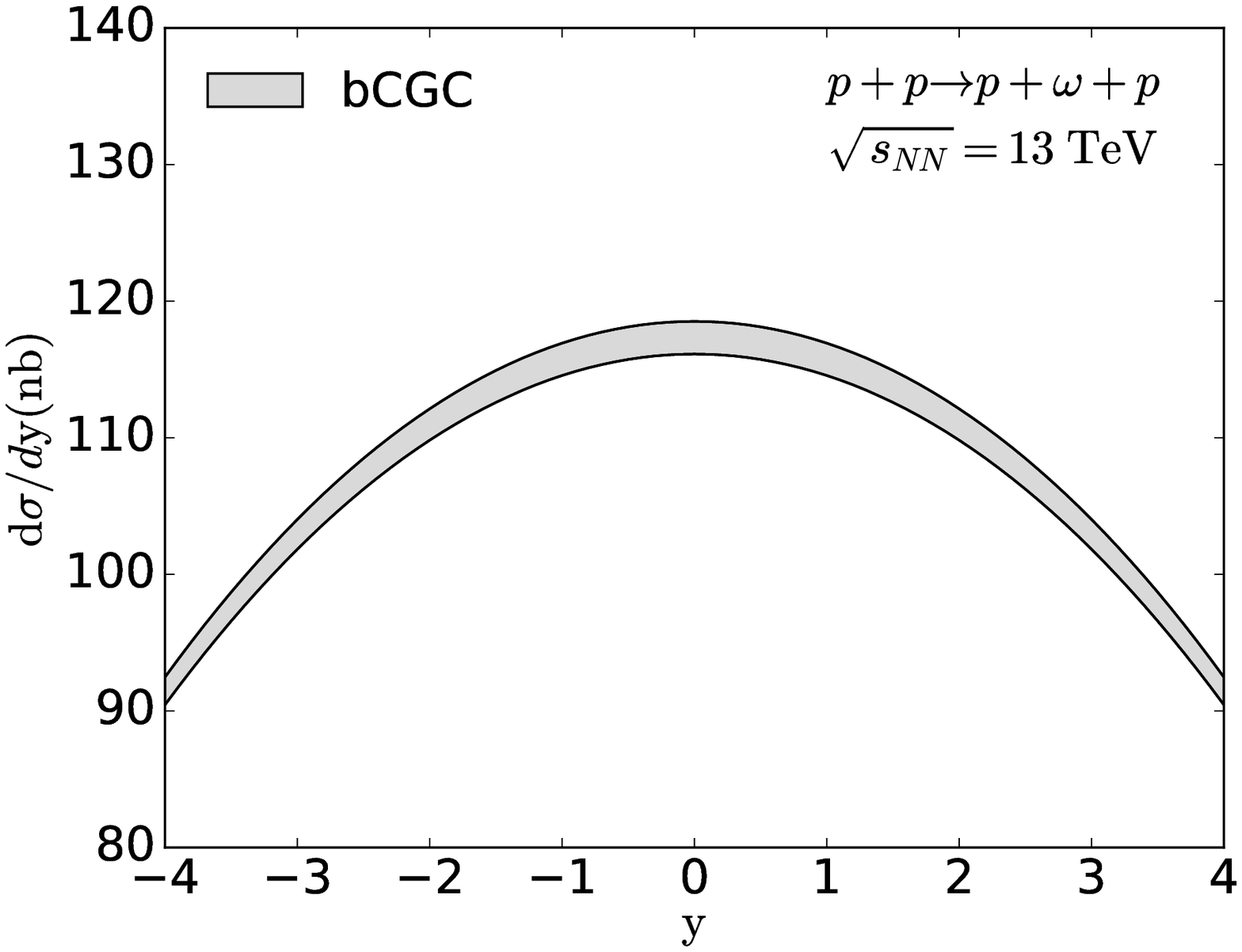}
  	\caption{(Color online)Predictions of rapidity distributions of $\omega$ meson computed in bCGC model using the Boosted Gaussian wave function in proton-proton ultraperipheral collisions at the LHC. The upper band of bCGC are using parameters of Fit 2 and the lower band of bCGC are using parameters of Fit 1. }
  	\label{omega}
  \end{figure}
Finally, the rapidity distributions of $\phi$ and $\omega$ mesons are also computed in the bCGC model with the Boosted Gaussian wave function in this paper. The predictions are shown in Fig.~\ref{phi} and \ref{omega}. The quark mass is $m_q=0.14$ GeV in the calculations and the upper band of bCGC are using parameters of Fit 2 and the lower band of bCGC are using parameters of Fit 1. Since there is no information of the rapidity gap survival factors for these two mesons now, the rapidity gap survival factors are taken as unity for $\phi$ and $\omega$ in the calculations. In Ref.\cite{Cisek:2010jk}, the authors presented the exclusive $\phi$ production in proton-proton UPCs at the LHC. The rapidity distributions of $\phi$ at LHC are smaller than the results in the paper since the different approaches are employed in the two paper. There is no experimental data for the $\phi$ and $\omega$  mesons in proton-proton UPCs at the LHC. We hope that the experimental data will be measured in the future. We can compare the theoretical prediction with the experimental data.\\
\section{conclusion}
In this paper, we have studied the exclusive photoproduction of $J/\psi$, $\psi (2s)$, $\phi$ and $\omega$  in proton-proton UPCs at the LHC.  The bCGC model and the Boosted Gaussian wave functions are employed in the calculation. The parameters of the bCGC model are refitted with the experimental data released in 2015. The theoretical predictions of $J/\psi$ and $\psi(2s) $ mesons rapidity distributions are evaluated in bCGC model and compared with the experimental data measured by the LHCb collaboration.  It can be seen that the predictions of bCGC model give a good description to the experimental data. It is concluded that the bCGC are successful phenomenological model for the small-x physics and the Boosted Gaussian wave functions are good candidates for the $J/\psi$ and $\psi(2s)$ mesons. The rapidity gap survival factor is important in the calculations multiplied together with the photon flux. The quark mass is sensitive to the exclusive vector mesons photoproduction in proton-proton UPCs. The rapidity distributions of $\phi$ and $\omega$ mesons are also performed in this paper. The predictions of the $\phi$ and $\omega$ mesons can be employed in future experiments.
\section{Acknowledgements}
We thank the useful discussions with M. V. T. Machado and V. P. Gon\c{c}alves. This work is supported in part by Key Research Program of Frontier Sciences,CAS (Grant No QYZDY-SSW-SLH006) and the National 973 project in China (No:~2014CB845406).


\begin{thebibliography}{99}
	\bibitem{Bertulani:2005ru}
	C.~A.~Bertulani, S.~R.~Klein and J.~Nystrand,
	Ann.\ Rev.\ Nucl.\ Part.\ Sci.,\  {\bf 55}: 271 (2005)
	[nucl-ex/0502005].
	\bibitem{Baltz:2007kq}
	A.~J.~Baltz, G.~Baur, D.~d'Enterria, L.~Frankfurt {\it et al.},
	Phys.\ Rept.,\  {\bf 458}:1 (2008)
	[arXiv:0706.3356 [nucl-ex]].
\bibitem{Chekanov:2002xi}
S.~Chekanov {\it et al.} [ZEUS Collaboration],
Eur.\ Phys.\ J.\ C, {\bf 24}: 345 (2002)
[hep-ex/0201043].
\bibitem{Chekanov:2004mw}
S.~Chekanov {\it et al.} [ZEUS Collaboration],
Nucl.\ Phys.\ B, {\bf 695}: 3 (2004)
[hep-ex/0404008].
\bibitem{Aktas:2005xu}
A.~Aktas {\it et al.} [H1 Collaboration],
Eur.\ Phys.\ J.\ C, {\bf 46}: 585 (2006)
[hep-ex/0510016].
\bibitem{Alexa:2013xxa}
C.~Alexa {\it et al.} [H1 Collaboration],
Eur.\ Phys.\ J.\ C, {\bf 73}(6): 2466 (2013)
[arXiv:1304.5162 [hep-ex]].

	\bibitem{Aaij:2013jxj}
	R.~Aaij {\it et al.} [LHCb Collaboration],
	J.\ Phys.\ G, {\bf 40}:045001 (2013)
	[arXiv:1301.7084 [hep-ex]].
	\bibitem{Aaij:2014iea}
	R.~Aaij {\it et al.} [LHCb Collaboration],
	J.\ Phys.\ G, {\bf 41}: 055002 (2014)
	[arXiv:1401.3288 [hep-ex]].
\bibitem{LHCb:2016oce}
The LHCb Collaboration [LHCb Collaboration],
LHCb-CONF-2016-007, CERN-LHCb-CONF-2016-007, oai:cds.cern.ch:2209532.

	\bibitem{Abelev:2012ba}
	B.~Abelev {\it et al.}  [ALICE Collaboration],
	Phys.\ Lett.\ B, {\bf 718}: 1273 (2013)
	[arXiv:1209.3715 [nucl-ex]].
	
	\bibitem{Abbas:2013oua}
	E.~Abbas {\it et al.}  [ALICE Collaboration],
	Eur.\ Phys.\ J.\ C, {\bf 73}(11): 2617 (2013)
	[arXiv:1305.1467 [nucl-ex]].
	
	\bibitem{TheALICE:2014dwa}
	B.~B.~Abelev {\it et al.}  [ALICE Collaboration],
	Phys.\ Rev.\ Lett.,\  {\bf 113},(23): 232504 (2014)
	[arXiv:1406.7819 [nucl-ex]].
	\bibitem{Adam:2015sia}
	J.~Adam {\it et al.} [ALICE Collaboration],
	Phys.\ Lett.\ B, {\bf 751}: 358 (2015)
	[arXiv:1508.05076 [nucl-ex]].
	\bibitem{Adam:2015gsa}
	J.~Adam {\it et al.} [ALICE Collaboration],
	JHEP, {\bf 1509}: 095 (2015)
	[arXiv:1503.09177 [nucl-ex]].


	\bibitem{Klein:1999qj}
	S.~Klein and J.~Nystrand,
	Phys.\ Rev.\ C, {\bf 60}: 014903 (1999)
	[hep-ph/9902259].
	S.~R.~Klein, J.~Nystrand and R.~Vogt,
	Eur.\ Phys.\ J.\ C, {\bf 21}: 563 (2001)
	[hep-ph/0005157].
	S.~R.~Klein, J.~Nystrand and R.~Vogt,
	Phys.\ Rev.\ C, {\bf 66}: 044906 (2002)
	[hep-ph/0206220].
	\bibitem{Frankfurt:2002sv}
	L.~Frankfurt, M.~Strikman and M.~Zhalov,
	Phys.\ Rev.\ C, {\bf 67}: 034901 (2003)
	[hep-ph/0210303].
	L.~Frankfurt, V.~Guzey, M.~Strikman and M.~Zhalov,
	Phys.\ Lett.\ B, {\bf 752}: 51 (2016)
	[arXiv:1506.07150 [hep-ph]].

	\bibitem{Goncalves:2005yr}
	V.~P.~Goncalves and M.~V.~T.~Machado,
	Eur.\ Phys.\ J.\ C, {\bf 40}: 519 (2005)
	[hep-ph/0501099].
	V.~P.~Goncalves and M.~V.~T.~Machado,
	Phys.\ Rev.\ C, {\bf 80}: 054901 (2009)
	[arXiv:0907.4123 [hep-ph]].
 V.~P.~Goncalves and M.~V.~T.~Machado,
 Phys.\ Rev.\ C, {\bf 84}: 011902 (2011)
 [arXiv:1106.3036 [hep-ph]].
 V.~P.~Gonçalves, B.~D.~Moreira and F.~S.~Navarra,
 Phys.\ Lett.\ B, {\bf 742}: 172 (2015)
 [arXiv:1408.1344 [hep-ph]].

 \bibitem{Ryskin:1992ui}
 M.~G.~Ryskin,
 Z.\ Phys.\ C, {\bf 57}: 89 (1993).
 doi:10.1007/BF01555742
A.~D.~Martin, C.~Nockles, M.~G.~Ryskin and T.~Teubner,
Phys.\ Lett.\ B, {\bf 662}: 252 (2008)
[arXiv:0709.4406 [hep-ph]].
	V.~Guzey and M.~Zhalov,
	JHEP, {\bf 1310}: 207 (2013)
	[arXiv:1307.4526 [hep-ph]].



	\bibitem{Adeluyi:2012ph}
	A.~Adeluyi and C.~A.~Bertulani,
	Phys.\ Rev.\ C, {\bf 85}: 044904 (2012)
	[arXiv:1201.0146 [nucl-th]].
	A.~Adeluyi and T.~Nguyen,
	Phys.\ Rev.\ C, {\bf 87}(2): 027901 (2013)
	[arXiv:1302.4288 [nucl-th]].

	\bibitem{Toll:2012mb}
	T.~Toll and T.~Ullrich,
	Phys.\ Rev.\ C, {\bf 87}(2): 024913 (2013)
	[arXiv:1211.3048 [hep-ph]].
	E.~Andrade-II, I.~González, A.~Deppman and C.~A.~Bertulani,
	Phys.\ Rev.\ C, {\bf 92}: 064903 (2015)
	[arXiv:1509.08701 [hep-ph]].
	T.~Lappi and H.~Mantysaari,
	Phys.\ Rev.\ C, {\bf 87}(3): 032201 (2013)
	[arXiv:1301.4095 [hep-ph]].
	T.~Lappi and H.~Mantysaari,
	Phys.\ Rev.\ C, {\bf 83}: 065202 (2011)
	[arXiv:1011.1988 [hep-ph]].
	
	\bibitem{Xie:2016ino}
	Y.~P.~Xie and X.~Chen,
	Eur.\ Phys.\ J.\ C, {\bf 76}(6): 316 (2016)
	[arXiv:1602.00937 [hep-ph]].
		Y.~P.~Xie and X.~Chen,
		Nucl.\ Phys.\ A, {\bf 957}: 477 (2017)
		[arXiv:1512.08105 [hep-ph]].
\bibitem{Xie:2017mil}
		Y.~P.~Xie and X.~Chen,
		Nucl.\ Phys.\ A, {\bf 959}: 56 (2017).
Y.~P.~Xie and X.~Chen,
Nucl.\ Phys.\ A {\bf 970}, 316 (2018).
doi:10.1016/j.nuclphysa.2017.12.003
\bibitem{Forshaw:2003ki}
J.~R.~Forshaw, R.~Sandapen and G.~Shaw,
Phys.\ Rev.\ D, {\bf 69}: 094013 (2004)
[hep-ph/0312172].
	\bibitem{GolecBiernat:1998js}
	K.~J.~Golec-Biernat and M.~Wusthoff,
	Phys.\ Rev.\ D, {\bf 59}: 014017 (1998)
	[hep-ph/9807513].
	K.~J.~Golec-Biernat and M.~Wusthoff,
	Phys.\ Rev.\ D, {\bf 60}: 114023 (1999)
	[hep-ph/9903358].
	\bibitem{Bartels:2002cj}
	J.~Bartels, K.~J.~Golec-Biernat and H.~Kowalski,
	Phys.\ Rev.\ D, {\bf 66}: 014001 (2002)
	[hep-ph/0203258].
	\bibitem{Iancu:2003ge}
	E.~Iancu, K.~Itakura and S.~Munier,
	Phys.\ Lett.\ B, {\bf 590}: 199 (2004)
	[hep-ph/0310338].
	\bibitem{Soyez:2007kg}
	G.~Soyez,
	Phys.\ Lett.\ B, {\bf 655}: 32 (2007)
	[arXiv:0705.3672 [hep-ph]].

\bibitem{Ahmady:2016ujw}
M.~Ahmady, R.~Sandapen and N.~Sharma,
Phys.\ Rev.\ D, {\bf 94}(7): 074018 (2016)
[arXiv:1605.07665 [hep-ph]].
	\bibitem{Kowalski:2003hm}
	H.~Kowalski and D.~Teaney,
	Phys.\ Rev.\ D, {\bf 68}: 114005 (2003)
	[hep-ph/0304189].
	\bibitem{Kowalski:2006hc}
	H.~Kowalski, L.~Motyka and G.~Watt,
	Phys.\ Rev.\ D, {\bf 74}: 074016 (2006)
	[hep-ph/0606272].
	\bibitem{Rezaeian:2012ji}
	A.~H.~Rezaeian, M.~Siddikov, M.~Van de Klundert and R.~Venugopalan,
	Phys.\ Rev.\ D, {\bf 87}(3): 034002 (2013)
	[arXiv:1212.2974].
	\bibitem{Watt:2007nr}
	G.~Watt and H.~Kowalski,
	Phys.\ Rev.\ D, {\bf 78}: 014016 (2008)
	[arXiv:0712.2670 [hep-ph]].
	\bibitem{Rezaeian:2013tka}
	A.~H.~Rezaeian and I.~Schmidt,
	Phys.\ Rev.\ D, {\bf 88}: 074016 (2013)
	[arXiv:1307.0825 [hep-ph]].

	\bibitem{Jones:2013pga}
	S.~P.~Jones, A.~D.~Martin, M.~G.~Ryskin and T.~Teubner,
	JHEP, {\bf 1311}: 085 (2013)
	[arXiv:1307.7099 [hep-ph]].
\bibitem{Jones:2013eda}
S.~P.~Jones, A.~D.~Martin, M.~G.~Ryskin and T.~Teubner,
J.\ Phys.\ G, {\bf 41}: 055009 (2014)
[arXiv:1312.6795 [hep-ph]].
\bibitem{Jones:2016icr}
S.~P.~Jones, A.~D.~Martin, M.~G.~Ryskin and T.~Teubner,
J.\ Phys.\ G, {\bf 44}(3): 03LT01 (2017)
[arXiv:1611.03711 [hep-ph]].
	\bibitem{Khoze:2013dha}
	V.~A.~Khoze, A.~D.~Martin and M.~G.~Ryskin,
	Eur.\ Phys.\ J.\ C, {\bf 73}: 2503 (2013)
	[arXiv:1306.2149 [hep-ph]].
\bibitem{Ducati:2013tva}
M.~B.~Gay Ducati, M.~T.~Griep and M.~V.~T.~Machado,
Phys.\ Rev.\ D, {\bf 88}: 017504 (2013)
[arXiv:1305.4611 [hep-ph]].
%
\bibitem{Khoze:2002dc}
V.~A.~Khoze, A.~D.~Martin and M.~G.~Ryskin,
Eur.\ Phys.\ J.\ C, {\bf 24}: 459 (2002)
[hep-ph/0201301].
	\bibitem{Shuvaev:1999ce}
	A.~G.~Shuvaev, K.~J.~Golec-Biernat, A.~D.~Martin {\it et al},
	Phys.\ Rev.\ D, {\bf 60}: 014015 (1999)
	[hep-ph/9902410].
	\bibitem{Armesto:2014sma}
	N.~Armesto and A.~H.~Rezaeian,
	Phys.\ Rev.\ D, {\bf 90}(5): 054003 (2014)
	[arXiv:1402.4831 [hep-ph]].
	
	\bibitem{Abramowicz:2015mha}
	H.~Abramowicz {\it et al.} [H1 and ZEUS Collaborations],
	Eur.\ Phys.\ J.\ C, {\bf 75}(12): 580 (2015)
	[arXiv:1506.06042 [hep-ex]].
	\bibitem{Aaron:2009aa}
	F.~D.~Aaron {\it et al.} [H1 and ZEUS Collaborations],
	JHEP, {\bf 1001}: 109 (2010)
	[arXiv:0911.0884 [hep-ex]].
	\bibitem{Abramowicz:1900rp}
	H.~Abramowicz {\it et al.} [H1 and ZEUS Collaborations],
	Eur.\ Phys.\ J.\ C, {\bf 73}(2): 2311 (2013)
	[arXiv:1211.1182 [hep-ex]].
\bibitem{Fiore:2014oha}
R.~Fiore, L.~Jenkovszky, V.~Libov and M.~Machado,
Theor.\ Math.\ Phys. ,\  {\bf 182}(1): 141 (2015)
[Teor.\ Mat.\ Fiz.,\ {\bf 182}(1): 171 (2014)]
[arXiv:1408.0530 [hep-ph]].	
\bibitem{Goncalves:2016sqy}
V.~P.~Goncalves, B.~D.~Moreira and F.~S.~Navarra,
Phys.\ Rev.\ D, {\bf 95} (5): 054011 (2017)
[arXiv:1612.06254 [hep-ph]].
\bibitem{Cisek:2010jk} 
A.~Cisek, W.~Schafer and A.~Szczurek,
Phys.\ Lett.\ B, {\bf 690}: 168 (2010)
[arXiv:1004.0070 [hep-ph]].
\end{thebibliography}
\end{document}